\documentclass[copyright]{eptcs}
\providecommand{\event}{TTC 2011} 

\usepackage{tikz}

\usepackage{listings}
\usepackage{graphicx}
\usepackage{xspace}
\usepackage{csquotes}
\usepackage{morefloats}

\newcommand{\origttfamily}{}
\let\origttfamily=\ttfamily
\renewcommand{\ttfamily}{\origttfamily \hyphenchar\font=`\-}

\newcommand{\name}[1]{{\sc #1}\xspace}
\newcommand{\grgen}{\name{GrGen}}
\newcommand{\grgennet}{\name{GrGen.NET}}
\newcommand{\grshell}{\name{GrShell}}

\lstdefinelanguage{grgen}
{morekeywords={actions,using,rule,test,pattern,replace,if,for,eval,negative,independent,node,edge,graph,modify,delete,class,model,connect,enum,abstract,const,exec,emit,alternative,multiple,iterated,optional,yield,copy,sequence,def,evalhere,emithere}, 
sensitive=true,
morecomment=[l]{//},
morecomment=[s]{/*}{*/},
morestring=[b]",
basicstyle=\ttfamily\footnotesize,
keywordstyle=\itshape,
tabsize=2,
}

\lstdefinelanguage{grshell}
{morekeywords={xgrs,debug,import,dump,add,set,node,edge,group,by,hidden,outgoing,labels,off,shortinfotag,exclude,shape,color,rhomb,white},
sensitive=true,
morecomment=[l]{\#},
morecomment=[s]{/*}{*/},
morestring=[b]",
basicstyle=\ttfamily\footnotesize
}

\lstnewenvironment{grgenlet}[1][]
{\lstset{language=grgen, numbers=none, tabsize=3, #1}}
{}

\title{Saying Hello World with GrGen.NET - A Solution to the TTC 2011 Instructive Case}
\author{Sebastian Buchwald \quad \quad Edgar Jakumeit
\institute{Karlsruhe Institute of Technology (KIT)}
\email{buchwald@kit.edu \quad \quad \phantom{~~~~~~~~~~~~~~~~}}
}
\def\titlerunning{Saying Hello World with GrGen.NET - A Solution to the TTC 2011 Instructive Case}
\def\authorrunning{Sebastian Buchwald and Edgar Jakumeit}
\begin{document}
\maketitle

\begin{abstract}
We introduce the graph transformation tool GrGen.NET (\url{www.grgen.net}) by solving
the Hello World Case \cite{helloworldcase} of the Transformation Tool Contest 2011
which consists of a collection of small transformation tasks;
for each task a section is given explaining our implementation.
\end{abstract}

\section{What is GrGen.NET?}

\grgennet\ is an application domain neutral graph rewrite system, the feature highlights regarding practical relevance are:
\begin{description}\itemsep -2pt
\item[Fully Featured Meta Model:] \grgennet\ uses attributed and typed multigraphs with multiple inheritance on node/edge types. Attributes may be typed with one of several basic types, user defined enums, or generic set, map, and array types.
\item[Expressive Rules, Fast Execution:] The expressive and easy to learn rule specification language allows for a  straightforward formulation of even complex problems, with an optimized implementation yielding high execution speed at modest memory consumption; outstanding features are iterated and recursive pattern matching and rewriting.
\item[Programmed Rule Application:] \grgennet\ supports a high-level rule application control language, Graph Rewrite Sequences (GRS), offering sequential, logical, iterative and recursive control plus variables and storages for the communication of processing locations between rules.
\item[Graphical Debugging:] \grshell, \grgennet's command line shell, offers interactive execution of rules, visualising together with yComp the current graph and the rewrite process. This way you can see what the graph looks like at a given step of a complex transformation and develop the next step accordingly. Or you can debug your rules and sequences.
\end{description}

\section{Hello World!}

The first task is to create a \texttt{Greeting} node with appropriate text.
To solve the task we use the \grgen rule from \texttt{HelloWorld.grg} shown below that creates the required graph when being executed:

\begin{lstlisting}[language=grgen]
rule createHelloWorld {
    replace {
        greeting:helloworld_Greeting;

        eval {
            greeting._text = "Hello World";
        }
    }
}
\end{lstlisting}

Rules in GrGen consist of a pattern part specifying the graph pattern to match and a nested rewrite part specifying the changes to be made.
The pattern part is built up of node and edge declarations or references with an intuitive syntax:
Nodes are declared by \texttt{n:t}, where \texttt{n} is an optional node identifier, and \texttt{t} its type.
An edge \texttt{e} with source \texttt{x} and target \texttt{y} is declared by \texttt{x -e:t-> y}, whereas \texttt{-->} introduces an anonymous edge of type \texttt{Edge}.
Nodes and edges are referenced outside their declaration by \texttt{n} and \texttt{-e->}, respectively.
Attribute conditions can be given within \texttt{if}-clauses.

The rewrite part is specified by a \texttt{replace} or \texttt{modify} block nested within the rule.
With \texttt{replace}-mode, graph elements which are referenced within the replace-block are kept, graph elements declared in the replace-block are created, and graph elements declared in the pattern, not referenced in the replace-part are deleted.
With \texttt{modify}-mode, all graph elements are kept, unless they are specified to be deleted within a \texttt{delete()}-statement.
Attribute recalculations can be given within an \texttt{eval}-statement.
These and the language elements we introduce later on are described in more detail in the GrGen.NET user manual \cite{GrGenUserManual};
a discussion of the merits of textual compared to visual languages can be found in \cite{sttt}.

The first rule shown above consists of an empty pattern part and a replace part that creates a new node \texttt{greeting} of type \texttt{helloworld\_Greeting} and \texttt{eval}uates the corresponding \texttt{\_text} attribute.
The creation rule for the extended metamodel given in \autoref{fig:helloWorldExt.grg} is similar in structure, just more voluminous.
A \texttt{Greeting} node is created and linked with a \texttt{greetingMessage} edge to a \texttt{GreetingMessage} node.
Furthermore, it is linked with a \texttt{person} edge to a \texttt{Person} node.
Then the attributes of the \texttt{message} and the \texttt{person} are initialized to the requested values (and the containment indices for XMI are set).

The files containing the rules seen so far start with a \texttt{using} statement
\verb#using helloworld__ecore;#
which imports the GrGen metamodel which was generated from the given Ecore metamodel.
GrGen.NET does not support importing Ecore metamodels directly (in contrast to GXL and native GRS files).
Instead we supply an import filter generating an equivalent GrGen-specific graph model (.gm file) from a given Ecore file
by mapping classes to GrGen node classes,
their attributes to corresponding GrGen attributes,
and their references to GrGen edge classes.
Inheritance is transferred one-to-one, and enumerations are mapped to GrGen enums.
Class names are prefixed by the names of the packages they are contained in to prevent name clashes;
the same holds for references which are prefixed by their node class name,
and node/edge attributes which are prefixed by an underscore.
This name mangling can be seen in the first rule given, in the following listings it was removed due to space constraints and for the sake of readability,
up to the migration chapter, from there on it is needed again to prevent type ambiguity.
We regard this name mangling import of Ecore as the only weak point of the solution.
Normally you are not interested in importing only a model, but you want to import an instance graph XMI (adhering to the Ecore model, thus adhering to the just generated equivalent GrGen graph model).
Such an instance graph is imported by the filter, too, serving as the host graph for the following transformations
(this can be seen in \autoref{fig:count.grs}, in this section we are only interested in the metamodel).
The import process described is realized by the \texttt{import} command of the GrShell.

\begin{figure}
\centering
\includegraphics[scale=0.2]{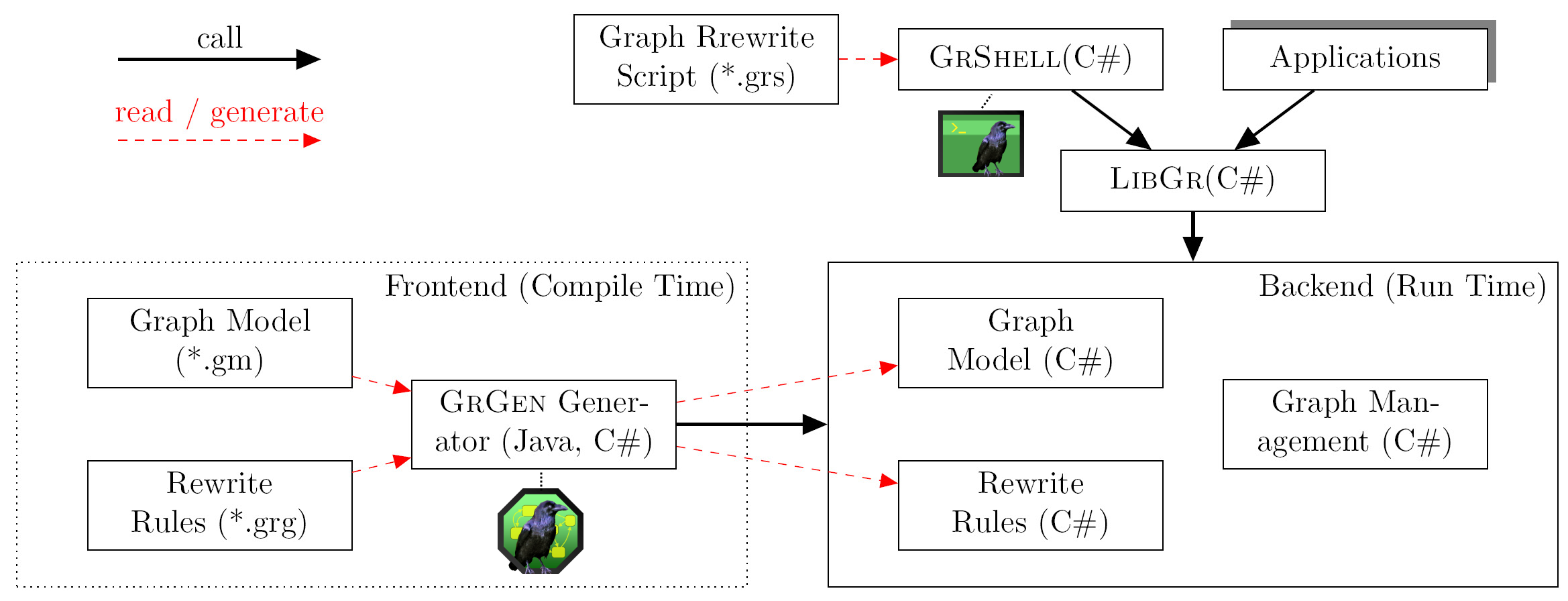}
\caption{The structure of the GrGen.NET system}
\label{fig:grgenoverview}
\end{figure}

The GrShell offers the host environment for applying the rules given in the \texttt{.grg} files.
The command line shell may be operated in interactive mode or in batch mode, for the tasks of this case we use it in batch mode by executing graph rewrite scripts, i.e. \texttt{.grs} files.
The high level workflow of using GrGen is shown in \autoref{fig:grgenoverview}: the user written model (\texttt{.gm}) and rule (\texttt{.grg}) files are compiled by GrGen into .NET assemblies;
together with a further runtime library these form the algorithmic core, to be accessed by user applications via the API offered by LibGr.
One such application is the GrShell, accompanied by the graph viewer yComp it offers a rapid prototyping environment for graph transformation.
For the first HelloWorld task the Shell executes the script given in \autoref{fig:helloWorld.grs}.
The first line imports the Ecore metamodel \texttt{helloworld.ecore} and loads the transformation rules declared in the rule file \texttt{HelloWorld.grg}.
The import process automatically generates the \grgen metamodel file \texttt{helloworld\_\_ecore.gm} that is used in the first rule and \autoref{fig:helloWorldExt.grg}.
We then execute an e\textbf xtended \textbf graph \textbf rewrite \textbf sequence that consists of one application of the \texttt{createHelloWorld} rule.
The graph rewrite sequences offer multiple operators for controlling rule execution and parameter passing between rules, several of them will be introduced later on, here we only execute a single parameterless rule once.
The third line is used to \texttt{show} the resulting graph with the yComp tool, and finally the shell execution is \texttt{quit}.

Let us continue with the model-to-text transformation, \autoref{fig:helloWorldToText.grg} shows the corresponding rule.
The rule matches a node of type \texttt{Greeting} and the corresponding \texttt{person} and \texttt{message}
(exactly what \autoref{fig:helloWorldExt.grg} created).
If a match is found, it creates a new node of type \texttt{StringResult} and assigns the concatenation of the \texttt{text} of the \texttt{message} and the \texttt{name} of the \texttt{person} to the \texttt{result} attribute.
The \texttt{StringResult} is then \texttt{emit}ted into an XMI file with the rule in \autoref{fig:Emitter.gri},
which gets \texttt{\#include}d into the main rule file.
The process is controlled by the shell script \autoref{fig:helloWorldExt.grs}.
The 5th line is used to \texttt{redirect} the output of the \texttt{emit} statements from stdout to the specified file.

\section{Count Matches with Certain Properties}

The next task is to count the number of occurrences of certain graph structures.
For each subtask the result needs to be wrapped in a node, this node is created by an application of the rule in \autoref{fig:createIntResult}.
The rule creates a node of type \texttt{IntResult}, initializes its \texttt{result} attribute to \texttt{0} and then \texttt{return}s the node out to the caller; it must be of type \texttt{IntResult} due to the output parameter declaration in the rule header with syntax \texttt{:(IntResult)}.
(Alternatively we could create the node in the shell with the \texttt{new} command which is the preferred way for creating non-trivial initial host graphs.)
The node with the count has to be written to an XMI file;
this is accomplished with a text emitting rule \texttt{emitInt} nearly identical to the already introduced one \texttt{emitString} available in the file \texttt{Emitter.grg}, you may have a look at the GrGen.NET SHARE image \cite{share} for details.

The first subtask consists of counting the number of nodes on the host graph.
This is achieved by using the \grgen rule shown in \autoref{fig:countNodes}.
The rule increments the \texttt{result} attribute of the \texttt{IntResult} parameter by one.
To get the count of all the nodes we execute the rule for all matches in the host graph---this
can be requested in the graph rewrite sequences calling the rules by enclosing the rule name in all-brackets,
as can be seen on line 4 of \autoref{fig:count.grs}.
Having a closer look at this line we see that the subtask is handled by the successive application of 3 rules.
The then-right operator \texttt{;>} executes the left sequence and then the right sequence, returning as result of execution the result of the execution of the right sequence.
The potential results of sequence execution are \emph{success} equaling \texttt{true} and \emph{failure} equaling \texttt{false}; a rule which matches counts as success.
The \texttt{IntResult} returned from the first rule is assigned to a variable \texttt{res}.
This variable is read before executing the second rule, its value is handed in as input argument to the second rule, in fact to all applications of the second rule.
The third rule emits (and deletes) the \texttt{IntResult} (it is not handed in, instead it gets matched in the \texttt{emitInt} rule).
(Alternatively we could count the number of nodes of a certain type \texttt{T} with \texttt{show num nodes T} in the \grshell.)

The count looping edges subtask is interesting because in the shell script calling it the keyword \texttt{debug} was prepended before the \texttt{xgrs} command (cf. line 6 of \autoref{fig:count.grs}). This causes sequence execution to start in debug mode, i.e. yComp is started visualizing the host graph and the rule matches of interest, and the sequence is executed stepwise under user control.
The situation right after the one match available for \texttt{countLoopingEdge} was found is displayed in \autoref{fig:countlooping} below,
the graph elements are annotated with the names of the pattern elements which matched them, cf. \autoref{fig:countLoopingEdges}.

\begin{figure}
\centering
\includegraphics[scale=0.4]{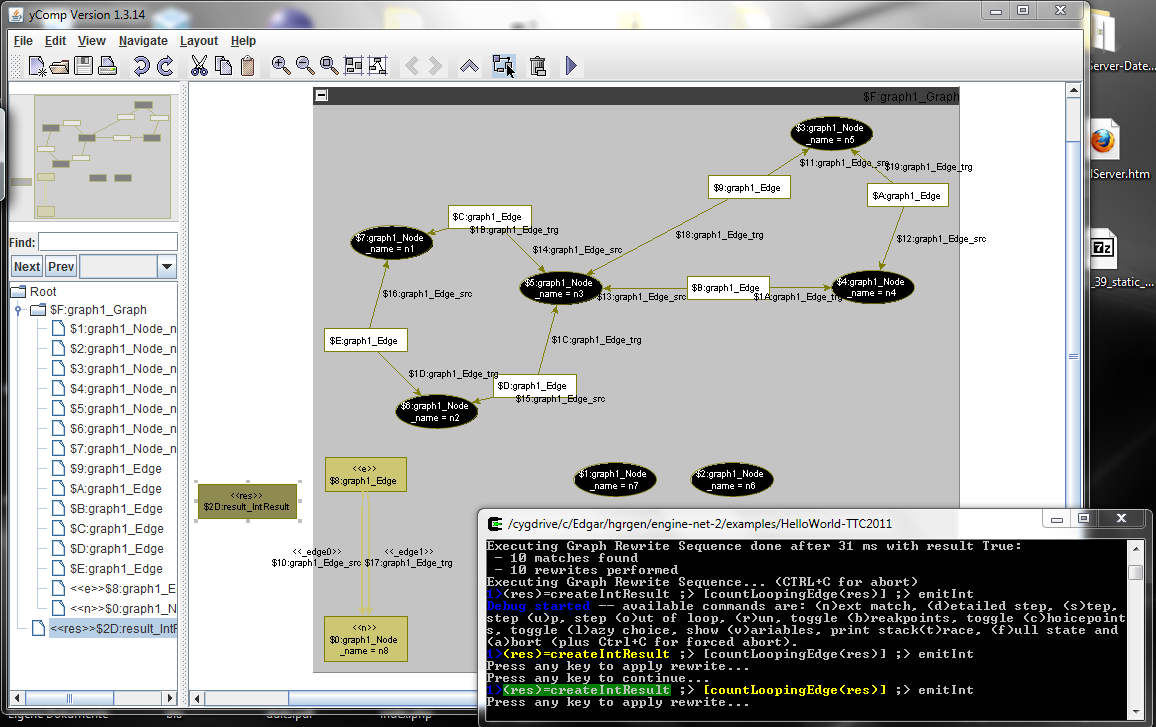}
\caption{Debugging the sequence calling countLoopingEdge}
\label{fig:countlooping}
\end{figure}

A major feature of yComp is its high configurability.
In line 2 of \autoref{fig:count.grs} we \texttt{include} a further shell script given in \autoref{fig:layout.grsi}, which is used to achieve the nice layout displayed in \autoref{fig:countlooping}.
You can use one of several available layout algorithms---with hierarchic and organic being the most useful ones, here we use circular.
You can configure for every available node or edge type in which color with what node shape or edge style it should be shown, with what attribute values or fixed text as element labels or tags it is to be displayed, or if it should be shown at all.
We used it to distinguish the \texttt{Node} nodes from the \texttt{Edge} nodes.
Furthermore you can configure graph nesting by registering edges at certain nodes to define a containment hierarchy, causing the nodes to become displayed as subgraphs containing the elements to which they are linked by the given edges.
This can be seen on lines 2 and 3 of \autoref{fig:layout.grsi} which cause all \texttt{Node} nodes and all \texttt{Edge} nodes to be contained in the graph node.

\autoref{fig:countDanglingEdges} shows the rule for the optional subtask of matching all dangling edges.
It matches an \texttt{Edge} node and then uses an \texttt{alternative} to match either the \texttt{missingSrc} pattern or the \texttt{missingTrg} pattern.
The \texttt{missingTrg} pattern matches the edge to the source node and uses a \texttt{negative} application condition (NAC) to ensure that the graph edge has no target node.
A \texttt{negative} pattern causes the matching of the enclosing pattern to fail if it is found in the graph.
Likewise, the \texttt{missingSrc} only matches an graph edge with a target node, but without a source node.
The rule \texttt{countIsolatedNode} in \autoref{fig:countIsolatedNode} is not matching as soon as one of the negatives is found.
The \texttt{countCycle} in \autoref{fig:countCycle} is a direct encoding of the specification, so there is no need to go into depth here.

\section{Reverse Edges}
To solve this task we need to reverse all edges.
The \grgen rule from \texttt{Reverse.grg} shown below employs retyping also known as relabeling on the \grgen edges to accomplish this task:

\begin{lstlisting}[language=grgen]
rule reverseEdge {
    alternative {
        Src {
            -src:src->;
            modify { -:trg<src>->; }
        }
        Trg {
            -trg:trg->;
            modify { -:src<trg>->; }
        }
    }
}
\end{lstlisting}

Retyping is specified with the syntax \texttt{y:t<x>}: this defines \texttt{y} to be a retyped version of the original node \texttt{x}, retyped to the new type \texttt{t};
for edges the syntax is \texttt{-y:t<x>->}.
After applying the rule with \texttt{xgrs [reverseEdge]} all source nodes are target nodes and all target nodes are source nodes.
This approach naturally reverses even dangling edges.

\section{Simple Migration}
To solve this task we need to migrate the graph from the graph metamodel used in the previous tasks to another graph metamodel which is characterized by introducing a superclass \texttt{GraphComponent} for \texttt{Node} nodes and \texttt{Edge} nodes.
Since the target metamodel has a similar structure, we simply use retyping as introduced in the previous section to migrate the graph, with the rules given in \autoref{fig:simpleToEvolved.grg} controlled by the sequence given in \autoref{fig:simpleToEvolved.grs}.
To keep things simple we offer an endogenous transformation, an exogenous one would be possible as well by matching all nodes, creating their counterparts, and writing traceability information to a storagemap, then matching all edges, creating their counterparts by looking up the correct nodes from the storagemap.
The name mangling from Ecore import is kept here (from now on) as removing it would render the types ambiguous.

In \autoref{fig:simpleToEvolved.grs} we apply each rule as long as possible (in each successive step searching one match then rewriting it, in contrast to the all bracketing introduced before which first collects all matches and then rewrites them at once).
This is denoted by the postfix star \texttt{*} causing the preceding sequence to be iterated as long as it succeeds.
The result of a star iteration is always success (in contrast to the plus \texttt{+} postfix which requires the preceding sequence to match at least once in order to succeed), so the complete sequence linked by strict conjunction operators \verb#&# succeeds (\texttt{true}), too. Disjunction \verb#|# is available as well, so are the lazy versions \verb#&&# and \verb#||# of the operators not executing the right sequence in case the result of the left sequence already determines the outcome.

The solution for the second (optional) target metamodel is similar to the first, so we only highlight the key difference: the second metamodel realizes edges by edges and not nodes anymore.
Thus model migration requires a non-isomorphic transformation step for edges,
here we use the rule shown in \autoref{fig:simpleToMoreEvolved.grg} and an additional rule to delete dangling edges.
Striving for perfection, we order the \texttt{linksTo} references by migrating the \texttt{graphEdge.index} and employing an additional fix-up rule that ensures that the index falls into the interval $0\ldots|\texttt{linksTo}|-1$.
Thus the (outgoing) edges are ordered the same way as they are ordered in the original graph.

\section{Delete Node with Specific Name and its Incident Edges}

Deleting a node with a given name is a trivial task, which can be seen in the rule given in \autoref{fig:deletetrivial.grg} using \texttt{modify} mode explicitly \texttt{delete}ing the matched node \texttt{if} it bears the name searched for.
The (optional) subtask of also deleting all incident edges is more interesting,
the rule from \texttt{Delete.grg} given below shows how this can be accomplished in \grgen:

\begin{lstlisting}[language=grgen]
rule deleteN1AndAllIncidentEdges {
    n:graph1_Node;
    if {n._name == "n1";}

    iterated {
        n <-- e:graph1_Edge;

        replace { }
    }
}
\end{lstlisting}

The rule above matches the node with the name \texttt{n1} and all incident edges in an \texttt{iterated} fashion.
The \texttt{iterated} construct munches eagerly the contained pattern as long as it is available in the graph and not yet matched;
it succeeds even if the contained pattern is not available in the graph, in contrast to the similar \texttt{multiple} construct which requires the pattern to be available at least once causing matching of the enclosing pattern to fail otherwise.
Since the \texttt{replace} parts are empty, all matched elements are deleted.
If edges were real edges and not nodes we could just delete the node, due to SPO-semantics the edges would be removed too.

\section{Insert Transitive Edges}
The last (optional) task is to compute $R\cup R^2$ from a graph representation of a relation $R$.
To solve this task we provide a rule \texttt{insertTransitiveEdge} given in \autoref{fig:transitivesimple.grg} that inserts a transitive edge if it is not available yet; we apply it on all matches by \texttt{xgrs [insertTransitiveEdge]}.
The edges can not be given directly in the pattern but must be enclosed in the positive application condition patterns denoted by the \texttt{independent}s in order to ensure that even for a multigraph with multiple edges between two nodes the transitive edges are not inserted multiple times.
The \texttt{hom(n1,n2,n3)} statement allows a homomorphic matching for the contained nodes,
i.e.\ they \emph{can} be matched to the same host graph node.
As a side remark: computing $R*$ would require nothing more than \texttt{xgrs insertTransitiveEdge*}
(then we could even move the edges out of the \texttt{independent}s; this was already the case in the first version of our solution for $R\cup R^2$ with the side effect of rendering it overly complicated, thanks to G\'{a}bor Bergmann for his hint to just employ an existantially quantified pattern, i.e. \texttt{independent}).
One could even compute the transitive closure only on demand in graph queries which require it by employing recursive patterns, 
similar to how they are used in our Program Understanding case solution \cite{programunderstanding}.

\section{Conclusion}
In this paper we presented a GrGen.NET solution to the Hello World!\ challenge of the Transformation Tool Contest 2011.
We were able to solve all tasks of the challenge, including the optional tasks,
introducing the reader alongside to a respectable amount of the functionality of GrGen.NET.

\nocite{*}
\bibliographystyle{eptcs}
\bibliography{refs}

\begin{thebibliography}{1}
\providecommand{\bibitemdeclare}[2]{}
\providecommand{\urlprefix}{Available at }
\providecommand{\url}[1]{\texttt{#1}}
\providecommand{\href}[2]{\texttt{#2}}
\providecommand{\urlalt}[2]{\href{#1}{#2}}
\providecommand{\doi}[1]{doi:\urlalt{http://dx.doi.org/#1}{#1}}
\providecommand{\bibinfo}[2]{#2}

\bibitemdeclare{misc}{GrGenUserManual}
\bibitem{GrGenUserManual}
\bibinfo{author}{Jakob Blomer}, \bibinfo{author}{Rubino Gei{\ss}} \&
  \bibinfo{author}{Edgar Jakumeit} (\bibinfo{year}{2011}):
  \emph{\bibinfo{title}{{The GrGen.NET User Manual}}}.
\newblock \bibinfo{howpublished}{\url{http://www.grgen.net}}.

\bibitemdeclare{misc}{share}
\bibitem{share}
\bibinfo{author}{Edgar Jakumeit} \& \bibinfo{author}{Sebastian Buchwald}
  (\bibinfo{year}{2011}): \emph{\bibinfo{title}{{SHARE} demo related to the
  paper Saying Hello World with {GrGen.NET} - A Solution to the {TTC} 2011
  Instructive Case}}.
\newblock
  \bibinfo{howpublished}{\url{http://is.ieis.tue.nl/staff/pvgorp/share/?page=C%
onfigureNewSession&vdi=XP-TUe_TTC11_GrGen_v2.vdi}}.

\bibitemdeclare{inproceedings}{programunderstanding}
\bibitem{programunderstanding}
\bibinfo{author}{Edgar Jakumeit} \& \bibinfo{author}{Sebastian Buchwald}
  (\bibinfo{year}{2011}): \emph{\bibinfo{title}{Solving the {TTC} 2011
  Reengineering Case with {GrGen}.{NET}}}.
\newblock In \bibinfo{editor}{\bibinfo{editor}{{Van Gorp}}} et~al.
  \cite{ttc2011eptcs}.

\bibitemdeclare{article}{sttt}
\bibitem{sttt}
\bibinfo{author}{Edgar Jakumeit}, \bibinfo{author}{Sebastian Buchwald} \&
  \bibinfo{author}{Moritz Kroll} (\bibinfo{year}{2010}):
  \emph{\bibinfo{title}{GrGen.NET}}.
\newblock {\sl \bibinfo{journal}{International Journal on Software Tools for
  Technology Transfer (STTT)}} \bibinfo{volume}{12}, pp.
  \bibinfo{pages}{263--271}, \doi{10.1007/s10009-010-0148-8}.

\bibitemdeclare{inproceedings}{helloworldcase}
\bibitem{helloworldcase}
\bibinfo{author}{Steffen Mazanek} (\bibinfo{year}{2011}):
  \emph{\bibinfo{title}{Hello World! An Instructive Case for the Transformation
  Tool Contest}}.
\newblock In \bibinfo{editor}{\bibinfo{editor}{{Van Gorp}}} et~al.
  \cite{ttc2011eptcs}.

\bibitemdeclare{proceedings}{ttc2011eptcs}
\bibitem{ttc2011eptcs}
\bibinfo{editor}{Pieter {Van Gorp}}, \bibinfo{editor}{Steffen Mazanek} \&
  \bibinfo{editor}{Louis Rose}, editors (\bibinfo{year}{2011}):
  \emph{\bibinfo{title}{TTC 2011: Fifth Transformation Tool Contest, Z\"urich,
  Switzerland, June 29-30 2011, Post-Proceedings}}.
  \bibinfo{publisher}{{EPTCS}}.

\end{thebibliography}

\vspace{2cm}
\appendix

\section{Code Listings}
\begin{figure}[ph]
\begin{lstlisting}[language=grgen]
rule createHelloWorldExt {
    replace {
        greeting:Greeting;
        greeting -e:greetingMessage-> message:GreetingMessage;
        greeting -f:person-> person:Person;

        eval {
            message.text = "Hello";
            person.name = "TTC Participants";

            e.index = 0;
            f.index = 0;
        }
    }
}
\end{lstlisting}
	\caption{HelloWorldExt.grg}
	\label{fig:helloWorldExt.grg}
\end{figure}

\begin{figure}[ph]
\begin{lstlisting}[language=grshell]
import helloworld.ecore HelloWorld.grg
xgrs createHelloWorld
show graph ycomp
quit
\end{lstlisting}
	\caption{HelloWorld.grs}
	\label{fig:helloWorld.grs}
\end{figure}

\begin{figure}[ph]
\begin{lstlisting}[language=grgen]
#include "Emitter.gri"

rule messageToText {
    greeting:Greeting;
    greeting -:greetingMessage-> message:GreetingMessage;
    greeting -:person-> person:Person;

    replace {
        result:StringResult;

        eval {
            result.result = message.text + " " + person.name + "!";
        }
    }
}
\end{lstlisting}
	\caption{HelloWorldExt.grg}
	\label{fig:helloWorldToText.grg}
\end{figure}

\begin{figure}[ph]
\begin{lstlisting}[language=grgen]
rule emitString {
    string:StringResult;

    replace {
        emit("<?xml version=\"1.0\" encoding=\"ASCII\"?>\n");
        // some further lines emitting XMI text not displayed
        emit("    result=\"" + string.result + "\"/>\n");
    }
}
\end{lstlisting}
	\caption{Emitter.gri}
	\label{fig:Emitter.gri}
\end{figure}
\begin{figure}[ph]
\begin{lstlisting}[language=grshell]
import helloworldext.ecore result.ecore HelloWorldExt.grg
xgrs createHelloWorldExt
show graph ycomp
xgrs messageToText
redirect emit "Model-to-text-grgen.xmi"
xgrs emitString
quit
\end{lstlisting}
	\caption{HelloWorldExt.grs}
	\label{fig:helloWorldExt.grs}
\end{figure}

\begin{figure}[ph]
\begin{lstlisting}[language=grgen]
rule createIntResult : (IntResult) {

    modify {
        res:IntResult;
        
        eval {
            res.result = 0;
        }
        
        return (res);
    }
}
\end{lstlisting}
	\caption{Count.grg}
	\label{fig:createIntResult}
\end{figure}

\begin{figure}[ph]
\begin{lstlisting}[language=grgen]
rule countNode(res:IntResult) {
    n:Node;

    modify {
        eval {
            res.result = res.result + 1;
        }
    }
}
\end{lstlisting}
	\caption{Count.grg}
	\label{fig:countNodes}
\end{figure}

\begin{figure}[ph]
\begin{lstlisting}[language=grshell]
import graph1.ecore result.ecore Graph1.xmi Count.grg
include layout.grsi
redirect emit "Count-nodes-grgen.xmi"
xgrs (res)=createIntResult ;> [countNode(res)] ;> emitInt
redirect emit "Count-looping-edges-grgen.xmi"
debug xgrs (res)=createIntResult ;> [countLoopingEdge(res)] ;> emitInt
# the other subtasks follow the two-line scheme given above
quit
\end{lstlisting}
	\caption{Count.grs}
	\label{fig:count.grs}
\end{figure}

\begin{figure}
\begin{lstlisting}[language=grgen]
rule countLoopingEdge(res:IntResult) {
    e:Edge -:src-> n:Node;
    e      -:trg-> n;

    modify {
        eval {
            res.result = res.result + 1;
        }
    }
}
\end{lstlisting}
	\caption{Count.grg}
	\label{fig:countLoopingEdges}
\end{figure}

\begin{figure}[ph]
\begin{lstlisting}[language=grshell]
debug set layout Circular
dump add node Graph group by hidden outgoing edges
dump add node Graph group by hidden outgoing nodes
dump set node Graph color lightgrey
dump set node Node shape circle
dump set node Node color black
dump set node Node textcolor white
dump add node Node infotag name
dump set node Edge color white
\end{lstlisting}
	\caption{layout.grsi}
	\label{fig:layout.grsi}
\end{figure}

\begin{figure}[ph]
\begin{lstlisting}[language=grgen]
rule countDanglingEdge(res:IntResult) {
    e:Edge;

    alternative {
        missingTrg {
            e -:src->;
            negative {
                e -:trg->;
            }

            modify { }
        }
        missingSrc {
            e -:trg->;
            negative {
                e -:src->;
            }

            modify { }
        }
    }

    modify {
        eval {
            res.result = res.result + 1;
        }
    }
}
\end{lstlisting}
	\caption{Count.grg}
	\label{fig:countDanglingEdges}
\end{figure}

\begin{figure}[ph]
\begin{lstlisting}[language=grgen]
rule countIsolatedNode(res:IntResult) {
    n:Node;
    negative {
        n <-:src-;
    }
    negative {
        n <-:trg-;
    }

    modify {
        eval {
            res.result = res.result + 1;
        }
    }
}
\end{lstlisting}
	\caption{Count.grg}
	\label{fig:countIsolatedNode}
\end{figure}

\begin{figure}[ph]
\begin{lstlisting}[language=grgen]
rule countCycle(res:IntResult) {
    n1:Node <-:src- :Edge -:trg-> n2;
    n2:Node <-:src- :Edge -:trg-> n3;
    n3:Node <-:src- :Edge -:trg-> n1;

    modify {
        eval {
            res.result = res.result + 1;
        }
    }
}
\end{lstlisting}
	\caption{Count.grg}
	\label{fig:countCycle}
\end{figure}

\begin{figure}[ph]
\begin{lstlisting}[language=grgen]
using graph1__ecore, graph2__ecore;

rule migrateGraph {
    n:graph1_Graph;
    modify {
        :graph2_Graph<n>;
    }
}
rule migrateNode {
    n:graph1_Node;
    modify {
        migrated:graph2_Node<n>;
        eval { migrated._text = n._name; }
    }
}
rule migrateEdge {
    e:graph1_Edge;
    modify {
        migrated:graph2_Edge<e>;
        eval { migrated._text = ""; }
    }
}
rule migrateEdgeSrc {
    -e:graph1_Edge_src->;
    modify {
        -migrated:graph2_Edge_src<e>->;
        eval { migrated.index = e.index; }
    }
}
rule migrateEdgeTrg {
    -e:graph1_Edge_trg->;
    modify {
        -migrated:graph2_Edge_trg<e>->;
        eval { migrated.index = e.index; }
    }
}
rule migrateGraphEdges {
    -e:graph1_Graph_edges->;
    modify {
        -migrated:graph2_Graph_gcs<e>->;
        eval { migrated.index = e.index; }
    }
}
rule migrateGraphNodes {
    -e:graph1_Graph_nodes->;
    modify {
        -migrated:graph2_Graph_gcs<e>->;
        eval { migrated.index = e.index; }
    }
}
\end{lstlisting}
	\caption{SimpleToEvolved.grg}
	\label{fig:simpleToEvolved.grg}
\end{figure}

\begin{figure}[ph]
\begin{lstlisting}[language=grgen]
xgrs migrateGraph* & migrateNode* & migrateEdge* & migrateEdgeSrc*\
       & migrateEdgeTrg* & migrateGraphEdges* & migrateGraphNodes*
\end{lstlisting}
	\caption{SimpleToEvolved.grs}
	\label{fig:simpleToEvolved.grs}
\end{figure}

\begin{figure}[ph]
\begin{lstlisting}[language=grgen]
rule migrateEdge {
    e:graph1_Edge;
    e <-graphEdge:graph1_Graph_edges-;
    e -:graph1_Edge_src-> src:Node;
    e -:graph1_Edge_trg-> trg:Node;

    modify {
        delete(e);

        src -migrated:graph3_Node_linksTo-> trg;

        eval {
            migrated.index = graphEdge.index;
        }
    }
}
\end{lstlisting}
	\caption{SimpleToMoreEvolved.grg}
	\label{fig:simpleToMoreEvolved.grg}
\end{figure}

\begin{figure}[ph]
\begin{lstlisting}[language=grgen]
rule deleteN1 {
    n:graph1_Node;

    if {n._name == "n1";}

    modify {
        delete(n);
    }
}
\end{lstlisting}
	\caption{Delete.grg}
	\label{fig:deletetrivial.grg}
\end{figure}

\begin{figure}[ph]
\begin{lstlisting}[language=grgen]
rule insertTransitiveEdge {
    n1:graph1_Node;
    n3:graph1_Node;
    hom(n1,n3);

    independent {
        hom(n1,n2,n3);
        n1 <-:graph1_Edge_src- :graph1_Edge -:graph1_Edge_trg-> n2:graph1_Node;
        n2 <-:graph1_Edge_src- :graph1_Edge -:graph1_Edge_trg-> n3;
    }
    negative {
        n1 <-:graph1_Edge_src- :graph1_Edge -:graph1_Edge_trg-> n3;
    }

    modify {
        n1 <-:graph1_Edge_src- :graph1_Edge -:graph1_Edge_trg-> n3;
    }
}
\end{lstlisting}
	\caption{TransitiveSimple.grg}
	\label{fig:transitivesimple.grg}
\end{figure}

\end{document}